# Molecular Bridge Mediated Ultra-Low-Power Gas Sensing

A. Banerjee, S. H Khan, S. Broadbent, A. Bulbul, K.H Kim,  R. Looper, C.H Mastrangelo and H. Kim,

**Abstract**

We report the electrical detection of captured gases through measurement of the tunneling characteristics of gas-mediated molecular junctions formed across nanogaps. The gas sensing nanogap device consists of a pair of vertically stacked gold electrodes separated by an insulating spacer of (~1.5 nm of sputtered α-Si and ~4.5 nm ALD $SiO_2$) which is notched ~10 nm along the edges of the top electrode. The exposed gold surface is functionalized with a self-assembled monolayer (SAM) of conjugated thiol linker molecules. When the device is exposed to a target gas (1, 5-diaminopentane), the SAM layer electrostatically captures the target gas molecules forming a molecular bridge across the nanogap. The gas capture lowers the barrier potential for electron tunneling from ~5 eV to ~0.9 eV across the notched edge regions and establishes additional conducting paths for charge transport between the gold electrodes, leading to a substantial decrease in junction resistance. We demonstrated an output resistance change of $> 10^8$ times on exposure to 80 ppm of diamine target gas as well as ultra-low standby power consumption of <15 pW, confirming electron tunneling through molecular bridges for ultra-low power gas sensing.

**Introduction**

The development of low-power Internet-of-Things (IoT) sensor-systems has been rigorously pursued by an increasing number of scientific communities to enable continuous access to various information around the globe. Detection of motion, humidity and temperature was proposed for the purpose of creating a user accessible database by the implementation of a low power IoT wireless sensor network by Laubhan [1]. Improvements to gas detection techniques have also been rigorously pursued to cover a wide range of commercial IoT applications including indoor air quality monitoring and for industrial applications such as detection of hazardous or combustible gases. Additionally, similar gas sensors have been deployed for remote and continuous environment monitoring to detect the presence of toxic gases such as volatile organic compounds (VOCs), CO, $SO_2$, $H_2S$ and $O_3$ [2].

Among such power-limited IoT applications, gas sensors remain as one of the most challenged components due to the power-hungry nature of gas sensing, especially in the required continuous operation mode. Most of the existing gas sensors consume relatively high-power (>10 μW) due to the requirements of elevated temperature or heating [3]. Recently, some low-power (sub-10 μW) gas sensors have been demonstrated, however, they displayed only limited output signal changes of less than few orders in magnitude, ultimately being limited in minimum detection amounts when utilizing ~μWatt level of power. These sensors achieved low power operation by utilizing self-heating under a bias voltage or by minimizing leakage electrical paths within the device, such as self-heating carbon nanotubes [4], Si-nanowires/$TiO_2$ core-shell heterojunctions [5], chemically gated Si nanowire/$SnO_2$ thin film FETs [6] and Pd based Si nano-membrane sensors [7]. They were limited in output signal range mainly because the transducer output varied linearly with analyte concentrations in a physically-connected device structure. For example, a CNT $NO_2$ sensor displayed only < 10 times change in the output current upon exposure to 0.9 ppm of target gas while consuming 0.8 μW of DC power, and an FET gas sensor produces an output of $< 10^2$ times at the exposure to 104 ppm of analyte while consuming ~ 1 μW DC power.

Such limited output signals can be further amplified without additional power consumption by physically forming an electrically conductive bridge across an otherwise-electrically isolated open space mediated by the capture of target gas molecules. This capture would lead to electron tunneling through the molecular bridge, resulting in multiple orders of magnitude current changes. The open air-gap could be realized by constructing a parallel plate structure between metal electrodes and matching the gap-distance with the molecular size of the target gas. The initial air-gap provides an extremely high DC off-resistance between electrodes. This leads to a minimal leakage current and hence a highly reduced standby power consumption. A similar nanogap concept was previously used but limited to in-liquid analysis and detection of bio-molecules such as proteins and DNA [8]–[14]. We previously introduced the concept for quantum tunneling based gas sensing with preliminary results [15]–[21].

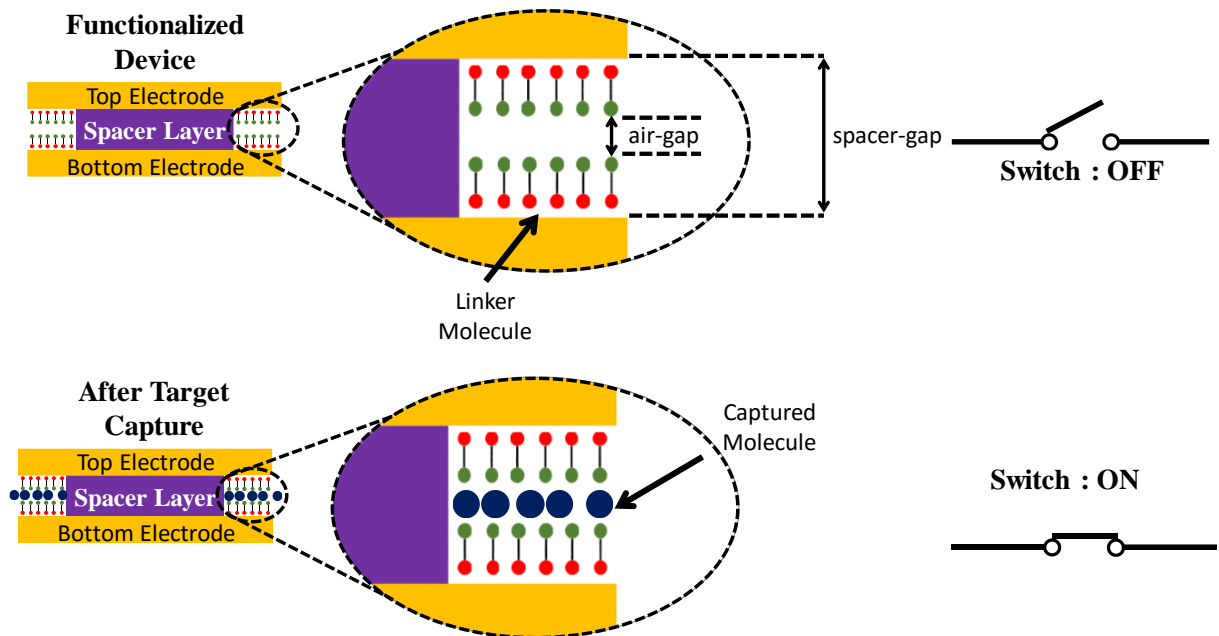

**Figure. 1.** Schematic of nanogap device after fabrication and analyte capture. Successful target capture turns the switch 'ON'.

This paper reports the detailed proof-of-concept results of a new family of ultra-low power nano air-gap sensors based on molecular bridge formation across nanogap electrodes. These air-gap sensors evidently demonstrated very high ON/OFF output signal ratios on account of molecular bridge mediated electron quantum tunneling. Specifically, this article reports the operation principle, device structure and fabrication, and sensor testing results.

**Tunneling Operation Based on Molecular Bridging**

Figure 1 illustrates the sensor's operating principle based on molecular bridging. The two electrodes are originally disconnected resulting in high tunneling resistance in the order of $10^9$ Ω as the 'OFF' state. These electrodes are coated with a self-assembled monolayer (SAM) of conjugated thiols which capture the target gas molecules. After molecular capture, the electrodes are physically connected through the formation of molecular bridges between them, thereby producing a path for enhanced quantum tunneling current and a significant reduction in junction resistance as the 'ON' state.

The electrical *I-V* characteristics and the device resistance is determined by the species trapped on the nanogap which can be two organic SAM layers separated by an air gap or a SAM-captured gas molecule-SAM chain. The electrical conductivity of these is examined below.

*Thiol SAM Capture Layer:* Generally speaking, the electrical properties of organic molecules can be determined by using AFM/STM [22], [23], MCB techniques [24] or can be simulated using computational chemistry methods [25]–[27]. Electrical transport through conventional alkane(di)thiol SAMs has been described as pure quantum tunneling across a thin dielectric film which has a rectangular potential barrier with image charge effects included, as described by J. Simmons in 1963 [28]. The dielectric constant of a typical alkane-thiol SAM layer has been determined by impedance measurements to be 2.1 according to Akkerman [29]. Akkerman also verified that the barrier height of alkane-thiol SAMs sandwiched between a pair of Au electrodes and protected by a layer of PEDOT:PSS (within the spacer stack), was in the range of 4-5 eV. Although alkane-thiols are poor conductors of electricity, the conductivity of thiol molecules can also be engineered by addition of alternative alkynes and aromatic rings within the molecule itself. This modification leads to a conjugated molecular arrangement which results in an amplified charge transport along the molecule on account of delocalized π-electron orbitals, which can freely move along the whole molecule. Bower also verified that conjugation in SAMs containing oligophenyl groups lead to a reduction in attenuation constant $\gamma$, which essentially meant an augmented conductance for molecules having a higher degree of conjugation [30]. This exponential dependence of resistance of SAMs on the SAM length ($R_{SAM} = k \cdot e^{(\gamma \cdot length)}$) has been extensively studied and verified. Using a combination of non-equilibrium Green's function, Density Functional Theory and confirmed I-V characteristics, Ratner [31] concluded that the degree of conjugation and overlap of orbital density of the molecular end group and the contact electrodes together determine the heightened conductivity of these molecular chains, on account of constant delocalized charge transport. Alkane di-amine molecules have also been studied using common computational chemistry algorithms [32] where the length dependence of electrical conductance has been observed to be exponential, similar to the thiol linker molecules. Generally, most organic molecules demonstrate an exponential dependence of electrical properties on molecule length. However, it must be noted that there is a transition molecular length beyond which the resistance dependence is almost ohmic in nature. Although this critical length depends on the specific organic molecule forming the molecular chain, generally for conjugated molecules, it is ~ 4.0 nm [33]. The exponential dependence of electrical resistance with length of the molecular chain provides a lower limit of the device resistance; hence for higher resistance change one should engineer the nanogap and SAM such that the SAM is as short as possible and the air gap should be the same length as the captured gas molecule. This combination maximizes the resistance ratio $R_{OFF}/R_{ON}$ and the dynamic range of the device. The electrical equivalent model for the device is discussed below.

*Electrical Equivalent Model:* The mechanism for electrical current conduction across the junction of our device is considered to be electron tunneling. In this device shown in the schematic of Figure 2 there are two possible electron conduction paths represented by current sources $I_S$ and $I_E(C_g)$, respectively. Current $I_S$ represents electrical conduction through the dielectric spacer stack under the overlapped region of the two electrodes. Current $I_E(C_g)$ represents electrical conduction through the molecular junction formed along the edges of the top electrode. This current is a function of the gas concentration $C_g$. In the absence of analyte target gas ($C_g = 0$), current mainly flows through the spacer stack in the overlap region under an overlap region of ~9 µm², thus $I_T \approx I_S$. On the other hand, when the target gas is captured between the nano-gaps along the edges, current ($I_E(C_g>0)$) will additionally flow through the newly formed molecular junction in the edge undercut area of ~0.06 µm². The device gap and linkers should be designed such that $I_E(C_g>0) \gg I_S$ and $I_T \approx I_E(C_g)$.

Total current flowing across the junction can be written as

$$I_T = I_S(V) + I_E(V, C_g) \tag{1}$$

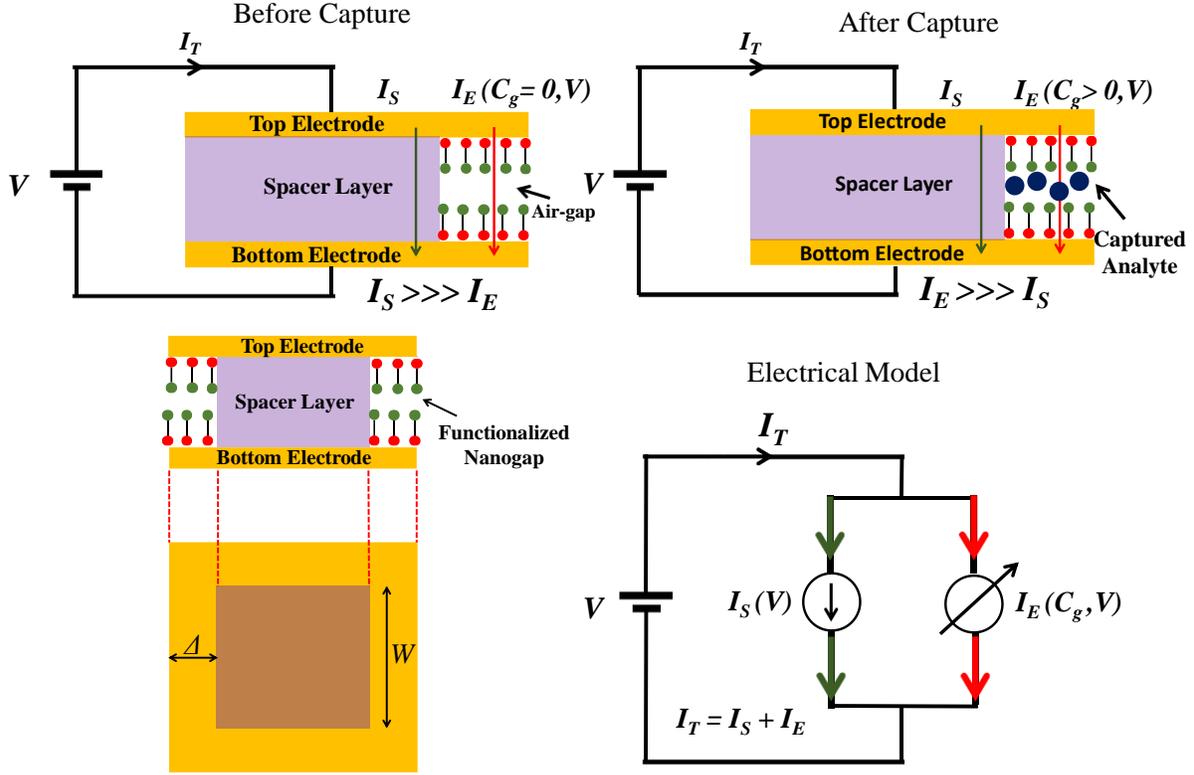

**Figure. 2** Schematic representation of current conduction before and after analyte gas capture and equivalent electrical model depicting the two current conducting paths.

where $I_T$ is the total current flowing across the nanogap sensor and $V$ is the biasing voltage across the device. The substrate and edge current components have different functional forms. The substrate component, which corresponds to conduction through the thin dielectric stacks, is readily modelled using the generalized Simmons' formula for tunneling current [28]. The current ($I_S$) across the overlap support region can be expressed as

$$I_S(V) = G_{SO} \cdot \left[\frac{\Phi_S}{q} \cdot e^{\left(-2 \cdot d_S \cdot \sqrt{\frac{2m}{\hbar^2} \cdot \Phi_S}\right)} + \left(\frac{\Phi_S}{q} + V\right) \cdot e^{\left(-2 \cdot d_S \cdot \sqrt{\frac{2m}{\hbar^2} \cdot (\Phi_S + qV)}\right)}\right] \quad (2)$$

where $\Phi_s$ is the barrier energy of the spacer dielectric, $d_s$ is the spacer layer thickness, $m$ is effective mass of tunneling electrons, $\hbar$ is the reduced Planck's constant, $q$ is the charge of electron. The parameter $G_{SO}=A_s \cdot g_{SO}$ is a conductance fitting parameter proportional to the area, $A_s=W^2$, of the spacer dielectric and $g_{SO}$ is a junction conductance factor.

The derivation of a mathematical expression for $I_E$ is considerably more difficult [34], [35], and in general it involves the calculation of non-equilibrium Green's functions which specify the ballistic electron transport across the various molecular levels of the trapped gas molecule. Such calculations can lead to tunneling resonances and negative resistance regimes. The model that we have adopted below is based on our experimental observations, which did not display any resonant tunneling characteristics.

In the non-tunneling case the *I-V* characteristics resemble those provided by a Simmons' –like model. However, when investigating current conduction across molecular chains, Ghosh [36] found that the Simmons' expression for tunneling current was unable to describe the *I-V* characteristics accurately. The reason was that certain assumptions made during the derivation of the Simmons' expression did not hold true for a molecular chain. N. Zimbovskaya [33] found that the conduction through molecular junctions can be modeled by combining a Wentzel-Kramer-Brillouin (WKB) approximation of the transmission coefficient and the Landauer formula which leads to a

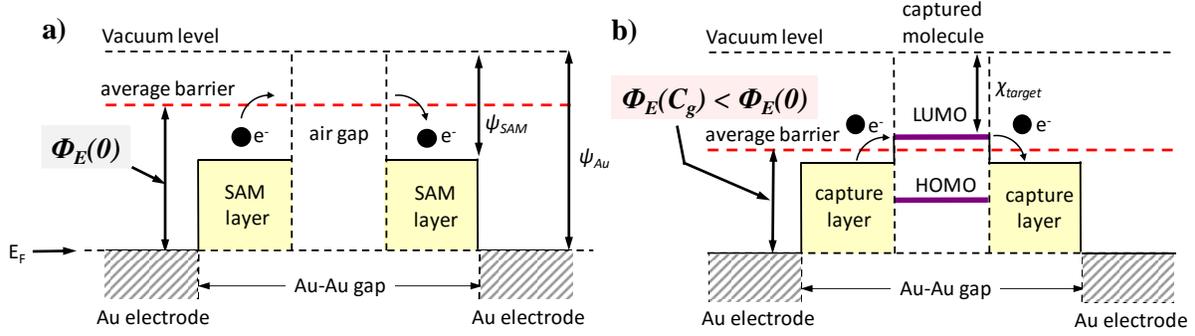

**Figure. 3 a)** Average edge energy barrier in absence of target analyte gas. **b)** Average edge energy barrier at a molecular junction established by the capture of a target gas molecule. The average energy barrier is lowered leading to a larger tunneling current.

mathematical expression of tunneling current similar to Simmons' formula. The tunneling current depends exponentially on a function of an average energy barrier as seen by the tunneling electrons. When the analyte gas is captured by our sensor, the dominant current is the gas-dependent edge current through the captured molecules and using Zimbovskaya formula it is given by:

$$I_E(V, C_G) = \left[\frac{\beta \cdot C_G}{1+ \beta \cdot C_G}\right] \cdot \Gamma_{EO} \cdot [\sqrt{\left(\frac{\Phi_E}{q} + \frac{V}{2}\right)} \cdot e^{-2 \cdot d_E \cdot \sqrt{\frac{2*m_E}{\hbar^2} \cdot \left(\Phi_E + \frac{qV}{2}\right)}} - \sqrt{\left(\frac{\Phi_E}{q} - \frac{V}{2}\right)} \cdot e^{-2 \cdot d_E \cdot \sqrt{\frac{2*m_E}{\hbar^2} \cdot \left(\Phi_E - \frac{qV}{2}\right)}}] \qquad (3)$$

where $\beta$ is a fitting parameter, $C_g$ is concentration of target analyte gas in the testing chamber, $\Phi_E$ is the average barrier potential of the hybrid molecular chain, and the distance $d_E = 2 \times$ (length of linker molecule) plus the length of analyte molecule. Two parameters are expected to change in this equation when a gas molecule is captured. The first and foremost important gas concentration dependent parameter is the average energy barrier $\Phi_E$ as shown in the zero-bias energy diagram of Figure 3. In Figure 3, $\Psi_{SAM}$ and $\Psi_{AU}$ are the work-functions of the SAM layer and the gold electrode respectively. $\chi_{target}$ is the electron affinity of the target gas.

Note that the average energy barrier depends on what is inside the edge gap. The energy barrier of the molecular bridge is obtained by fitting and comparing curves at different gas concentration with the Zimbovskaya formula to get an estimate of the average that electrons encounter. If there is no molecule captured, the average barrier is at the maximum given by the work function of the linker. If a molecule is captured, the average barrier established on that junction depends on the location of the HOMO and LUMO energy levels of the captured molecule. Curve-fitting and parameter extraction reveals that in general, the capture results in a lower average energy barrier $\Phi_E$ and hence a large enhancement of the device current.

The second concentration dependent factor is the leading term:

$$\left[\frac{\beta * C_G}{1+ \beta * C_G}\right] \Gamma_{EO} \qquad (4)$$

that is representative of the number of molecular junctions formed and is an indicative of a saturation-type Langmuir adsorption characteristic [31]. The parameter $\beta$, is a fitting parameter representing the adsorption characteristic, has a unit of 1/gas concentration and determines the low-concentration limit behavior of the edge current]. This equation implies that higher the surface concentration ($C_G$ of the target gas) is, the larger the number of molecular junctions formed $(\beta * C_G)/(1 + \beta * C_G)$ is, ultimately leading to the higher the edge current $I_E(V,C_G)$. The edge current saturates when all possible absorption sites are full. The parameter $\Gamma_{E0} = p \cdot g_{EO}$ is a fitting parameter proportional to the edge perimeter $p=4 \cdot (W+2\Delta)$. This parameter specifies the magnitude of the nonlinear edge conductance with a unit of $A \cdot V^{-1/2}$. Note that Eq. (3) also displays an exponential dependence of the molecular junction resistance on total molecular lengths which is in agreement with observations made by others [26][30][32].

*Device Design and Fabrication*

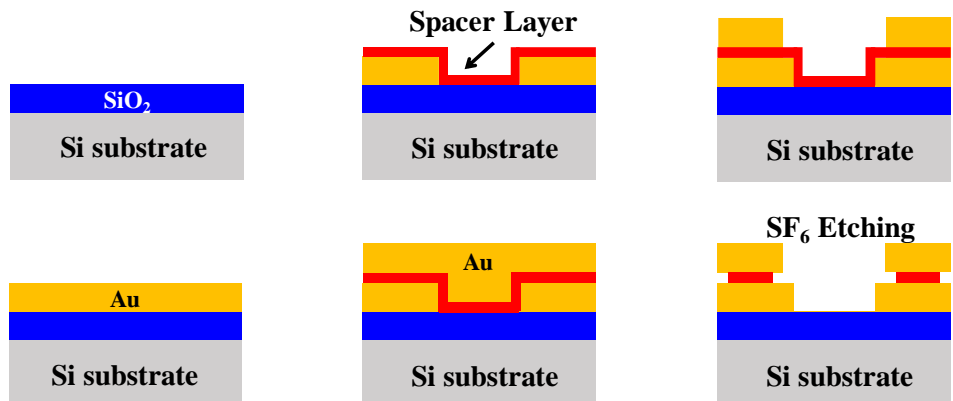

**Figure. 4** Simplified fabrication flow of nanogap electrodes.

Figure 4 shows the fabrication process. Details of the device fabrication method have been extensively discussed in our previous article [18]. The fabrication process started by growing 500 nm of thermal oxide on a Si wafer. Then, 20 nm of Cr and 200 nm of Au films were deposited and patterned to define the lower electrodes. Next, a $SiO_2$ layer (~4.5 nm) was deposited using plasma enhanced ALD at 200° C chuck temperature and an additional α-Si layer (~1.5 nm) was deposited, which together determined the thickness of the spacer stack. The thickness of the stack layers (ALD $SiO_2$ and DC sputtered α-Si) was verified using ellipsometry (Woollam Variable Angle Spectroscopic Ellipsometer (VASE)). On top of these spacer layers, the upper Au electrode layer (200 nm) was sputtered and then subsequently patterned using standard lithographic techniques. Finally, the space stack layers were etched away through $SF_6$ plasma dry etching, thereby forming an air-gap along the edges of the top electrode. Figure 5a shows the SEMs of the fabricated device where the overlap area was reduced to suppress parasitic current flows. The device footprint was 0.36 $mm^2$ and the overlap area was ~16 $\mu m^2$ (Figure 5-middle). Figure 5b confirmed the nanometer-scale air-gap formed between the upper and lower electrodes by high-resolution SEM imaging.

*Target Molecule and SAM Capture Linker Structure and Synthesis*

To demonstrate the validity of the proposed sensing mechanism, we aimed to detect a vapor of 1,5-pentanediamine, commonly known as cadaverine, as the target gas molecule. Figure 6 shows the chemical structure of target and linker molecules. The thiol linker molecule is comprised of three components: the sulfur head group, the aromatic and alkyl spacer chain and the benzoic acid tail group. The head group, consisting of thiol molecules (-SH), is capable of covalently bonding to the gold surface. The spacer chain defines the length of the linker molecule and due to its conjugated nature, allows for augmented electron flow throughout the molecule. The capture of our targets is achieved by the tail segment of the linker, which in this case is composed of benzoic acid. The terminal carboxylic acid from the benzoic acid, the matching length of the spacer chain to the nanogap distance together determine the specificity of the target gas molecules that are to be captured.

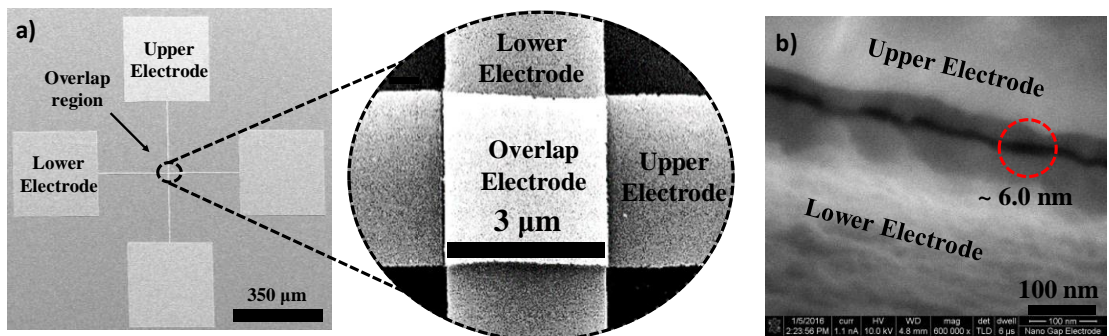

**Figure 5**. **a)** Schematic of electrode layout for nanogap device. **b)** SEM image of fabricated device

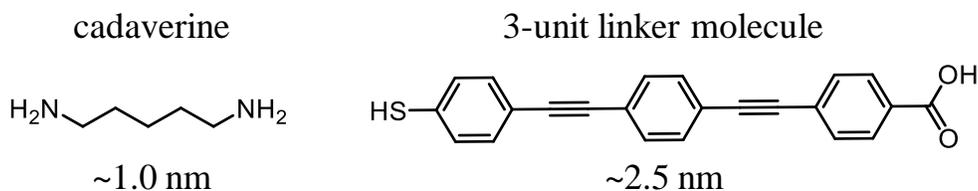

**Figure. 6** Chemical structure and molecule-length of the target molecule (on the left) and linker molecule (on the right). The IUPAC name of the linker molecule is (4-((4-((4-mercaptophenyl)ethynyl)phenyl)ethynyl)benzoic acid).

*Capture Linker Synthesis*: The linkers, consisting of three benzene rings (Figure 6-right), were synthesized to be able to capture amine group molecules (Figure 6-left). Figure 7 shows the chemical synthesis process starting from commercially available 4-iodobenzoate (1) that first underwent a Sonogashira reaction with trimethylsilyl acetylene to introduce the first alkyne (2). Next, the alkyne was freed by removing the protecting group (3) and subsequently went through another Sonogashira with 1-Bromo-4-Iodobenzene to introduce the second phenyl ring on the free alkyne (4). Identical steps were repeated to add the second alkyne to the second phenyl ring and to remove the protecting group from the alkyne to yield the free alkyne terminal (5, 6). As the coupling partner to the free alkyne terminal, commercially available 4-iodobenzenesulfonyl chloride (7) was reduced to the sulfonic acid and converted to the corresponding thioacetate (8) in order to prevent a free thiol from poisoning the palladium in the final Sonogashira coupling. Then, these two compounds were coupled via a final Sonogashira reaction to provide the fully conjugated ter phenyl linker (9). The thioacetate part of the coupled compounds were cleaved through saponification using lithium hydroxide to yield the final linker molecule. During the process the methyl ester was converted to the corresponding carboxylic acid that was designed to capture the target amine group molecules (10). The thiol end-group of the molecule covalently bound to the exposed gold surface of the electrodes, forming an Au-S bond as a firm linker coating.

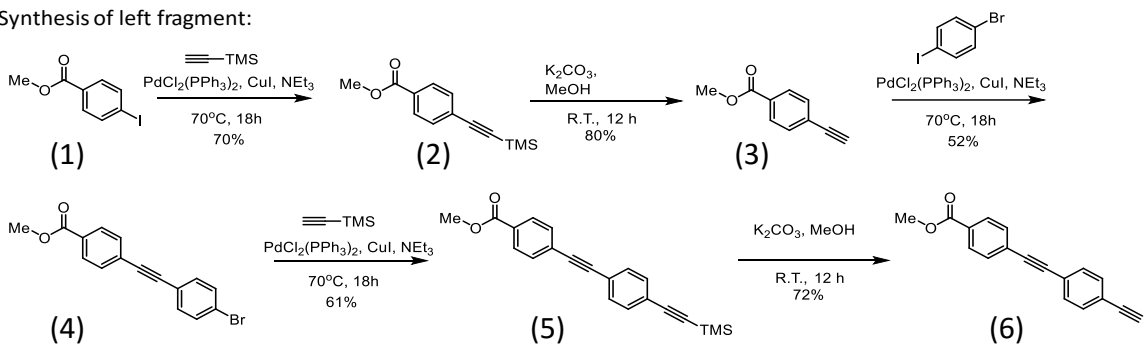

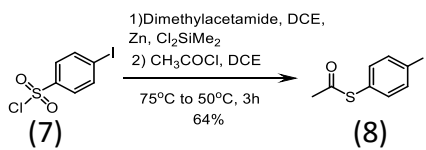

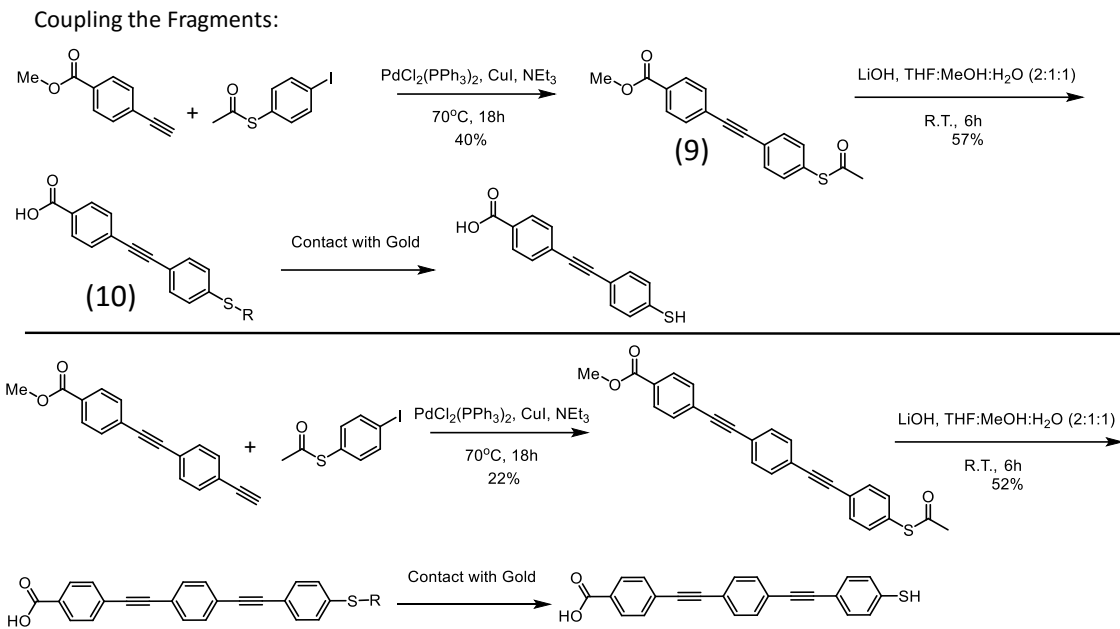

**Figure. 7 a)** Synthesis of left and right fragments of linker molecule **b)** coupling of the left and right fragments to form the thiol-linker molecule.

*SAM coating procedure, coating characterization and target capture specifics, and linker conductivity measurements*

Thiol-SAM chemisorption on gold surfaces, first reported by Bell Laboratories in 1983 [37], was performed by dissolving the linker chemical in a solution of dimethyl sulfoxide (DMSO) and ethanol, and immersing the gold coated device for a specified amount of time. In order to determine the adequate time, different immersion periods of 12, 24 or 48 hours were applied while the fluorescence imaging of the captured cadaverine target were compared. Multiple glass samples (1×1 cm$^2$) with pre-deposited Cr/Au layers of 20/200 nm thickness were kept immersed for 12, 24 or 48 hours, respectively. They were then exposed to fluorescent liquid cadaverine (Alexa Fluor 488, Sigma Aldrich) for 2 minutes and followed by DI water wash and N$_2$ dry processes for surface inspection under a fluorescence microscope. Figures 8 a-c show fluorescence imaging of the tested samples after cadaverine capture process. The lighter spots in the image indicate successful capture of the fluorescent cadaverine molecule. Since the cadaverine is captured only by the linker molecule, the lighter spots are actually an indirect indication of successful chemisorption of the linker molecule on the surface of the gold sample. Fluorescent imaging showed that a longer immersion period resulted in higher densities of cadaverine-to-linker capture, resulting in 48 hour coating as our standard. The maximum immersion hours were limited to 48 hours due to metal peeling-off issues.

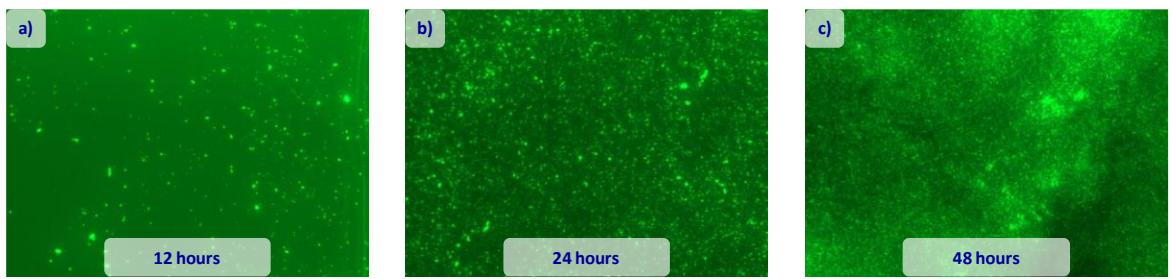

**Figure. 8** Fluorescence microscopy images of captured fluorescent cadaverine on gold samples immersed in a linker solution for **a)** 12 hours **b)** 24 hours **c)** 48 hours.

The coated linkers, without the target gas molecules, qualitatively exhibited augmented conductivity in comparison to commercially available non-conducting alkane-thiol linker molecules, as shown in Figure 9 a,b. The conductivity was measured using Peak Force Tunneling Atomic Force Microscopy (PF – TUNA, Bruker).

*Methods*

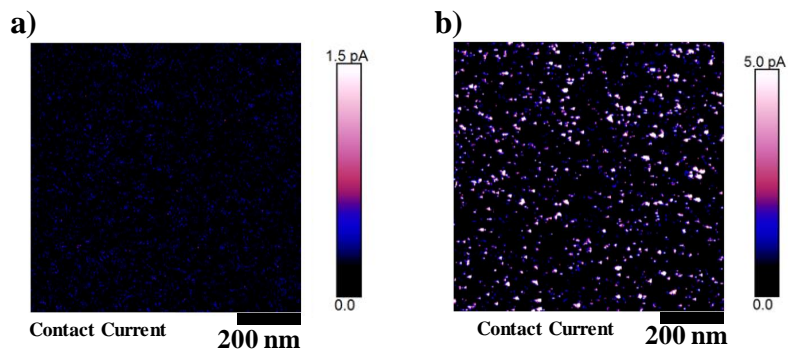

**Figure. 9** PF TUNA measurements of **(a)** commercially available non-conductive SAM of linker **(b)** synthesized conjugated SAM of linker

Gas concentration estimation. Concentrations of the target gas, cadaverine, was estimated by first establishing a calibration curve by injecting known sample volumes and comparing the signal intensities (areas) from the flame ionization detector (FID) and then comparing the unknown signal intensities to the calibration curve. For calibration, the utilized sample volumes were 0.1, 0.2, 0.3, 0.4 and 0.5 µL of 1% cadaverine diluted in water corresponding to a weight range of 6.2-31.2 ng. The calibration process produced a calibration equation between signal area (A) and injected mass (m): $A_{FID} = 31.18 \times m_{inj} - 0.03794$. Utilizing the equation, unknown concentrations were estimated. For example, an injection of 100 µL was calculated into $4.09 \times 10^{-6}$ mole at 250 °C and 1 atm, which and 100 µL is the injected gas volume. Assuming that all cadaverine evaporates, calculated moles of cadaverine in vapor phase would become 0.0013 µg/102.178 g/mole =$12.31 \times 10^{-12}$ mole. Therefore the PPM was $1.23 \times 10^{-11}/4.09 \times 10^{-6} \times 10^{6} = 3.01$ PPM.

Sensor Measurements in test chamber. The testing setup utilized a stainless steel cylindrical structure (height: 15 cm and diameter: 5.5 cm) with an injection hole sealed with a rubber septum seal and two electrical feedthroughs. The two electrical feedthroughs connected the fabricated sensor located inside the chamber to an impedance analyzer, Keithley 4200A-SCS, outside. The chip was supplied with a DC bias of 0.7 V across the nanogap-separated electrodes.

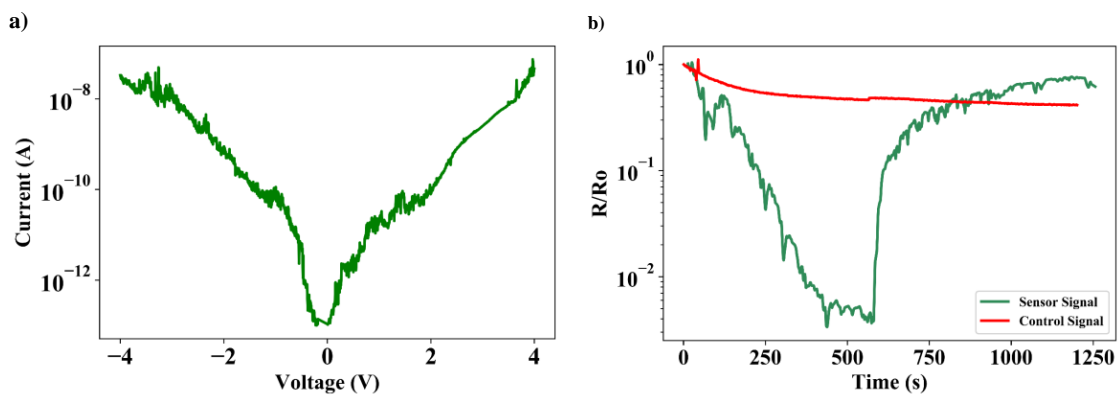

**Figure. 10 (a)** I-V curve of the completed device without gas exposure showing complete electrical isolation between upper and lower gold electrodes. **(b)** Sensor analysis over one complete cycle of 40 ppm target gas exposure and its removal. The sensor signal is compared to the response of a 'control' chip which is our nanogap device in absence of proper functionalization.

## Testing Results

### Experimental *I-V* Characteristics under gas exposure

The fabricated sensor, without being exposed to target gas, showed that the leakage current across the tunneling junction was minimal in the order of pA within a bias range of -4V and 4V, which resulted in an average DC resistance of 74 GΩ, as shown in Figure 10(a). This result confirmed the 'OFF' state of the device before the gas exposure, consuming ultra-low-power of pWs in the standby condition. When the fabricated sensor was exposed to a target gas at a concentration of 40 ppm, it clearly exhibited a sudden reduction of junction resistance of three orders of magnitude (Figure 10-(b)), clearly demonstrating switching to the 'ON' state. These results confirmed the physical disconnection and connection of an electrical path by molecular bridging, enabling high ON/OFF signal ratios as desired. The plot also shows control (reference) chip signal (a device which was not coated with the linker molecule and exposed to similar ppm amounts of target cadaverine gas). The feeble response of the control chip shows that in absence of proper functionalization, even after exposure to cadaverine gas, there is almost no sensor response as the thiol molecules are unable to capture the analyte molecules and the molecular switch is remains 'OFF'. Essentially, the sensor response from the control chip proves the validity of the linker molecules and their essential role in capturing the target gas molecules.

### Experimental and Fitted Model *I-V* Characteristics at Different Gas Concentrations

Figure 11 shows both experimental and model-predicting I-V plots at various concentrations and their tight agreement. As evident from the plot, exposure to ~80 ppm of cadaverine lead to a reduction of junction resistance by ~8 orders of magnitude. Theoretical parameter extraction predicted that the exposure of the fabricated device to ~80 ppm of cadaverine gas would reduce the average potential barrier from ~5 eV to ~0.9 eV, resulting in exponential tunneling current variation with the square-root of the barrier potential changes and that the fitting parameter $\beta$ and $\Gamma_{E0}$ would be 21.685/ppm and 0.3688 A/V$^{-1/2}$, respectively, in Equation (3). The maximum root-mean-square-error of the curve-fitting plots shown in Figure 11 was found to be ~ 45%, 1% and 0.6% of the average experimental data for the *I-V* characteristics of the nanogap junction after exposure to 0 ppm, 60 ppm and 80 ppm of cadaverine respectively. It must be noted that the predicted reduction in the average potential barrier was not a quantitative measurement of the HOMO-LUMO levels of individual target/linker molecules but a cumulative indication of the average potential barrier faced by tunneling electrons. In fact Zimbovskaya [33] mentions that the inherent disadvantage of this mathematical model is that although one can accurately describe *I-V* characteristics using Equation 3, detailed information including the electrostatic potential profile of the transport channel cannot be obtained. This is because of the WKB approximation for electron transmission functions used to derive the model in the first place.

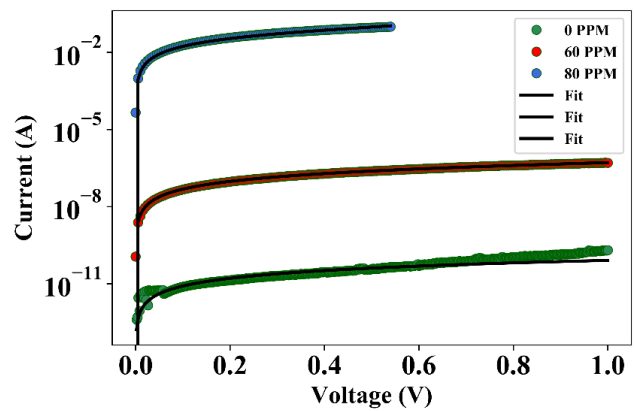

**Figure. 11** *I-V* measurements of the nanogap device after successful capture of cadaverine molecules at various concentrations of analyte.

**Selectivity of Gas Response**

In order to investigate the cross-sensitivity of the sensor, we exposed the device to analytes having different molecular end-groups, commonly found VOC's and gases such as Helium, Hydrogen and $CO_2$. We define the sensor response as the normalized junction resistance drop at steady-state after exposure to the analyte. Figure 12(a) shows the sensor response when exposed to these analytes as compared to that when exposed to the intended target gas, cadaverine. Figure 12(b) shows the response of the nanogap sensor when exposed to analytes having different molecular end-groups. Measurements reveal a $R_{OFF}/R_{ON}$ ratio of more than 3 orders of magnitude when exposed to only 40 ppm concentration of cadaverine whereas a maximum $R_{OFF}/R_{ON}$ ratio of ~1 order of magnitude, when exposed to much greater concentrations of the other analytes. The concentration of the VOCs was maintained at greater than 1000 ppm. To measure the device response in presence of other gases, we flooded the test chamber with the specific test gas and then monitored the resistance drop of the sensor. These results suggest a highly selective sensor action against most commonly found VOCs.

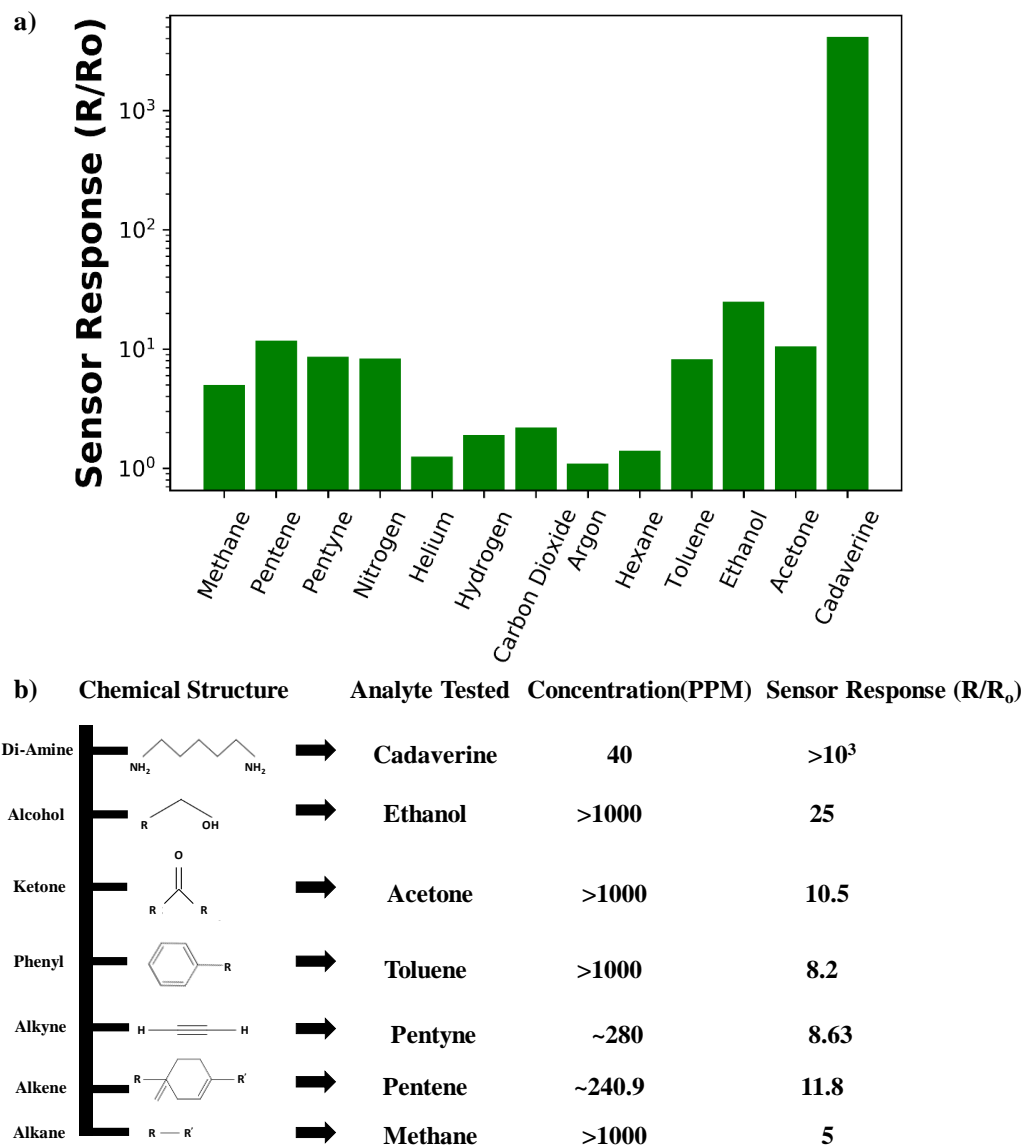

| | Chemical Structure | Analyte Tested | Concentration(PPM) | Sensor Response (R/R$_o$) |
|---|---|---|---|---|
| Di-Amine | | Cadaverine | 40 | >10$^3$ |
| Alcohol | | Ethanol | >1000 | 25 |
| Ketone | | Acetone | >1000 | 10.5 |
| Phenyl | | Toluene | >1000 | 8.2 |
| Alkyne | | Pentyne | ~280 | 8.63 |
| Alkene | | Pentene | ~240.9 | 11.8 |
| Alkane | | Methane | >1000 | 5 |

**Figure. 12 (a, b)** Device response to commonly found gases and analytes with different molecular end-groups.

*Transient Response Characteristics: Adsorption-Desorption dynamics*

Measurement results indicated that the dynamics of a junction resistance drop was governed by a quasi-irreversible process of cadaverine molecules forming hydrogen-bonds with the linker end group and forming a hybrid molecular bridge, as shown in Figure 13. Such a typical chemisorption process, ideally consisting of complete breaking of hydrogen bonds between the target gas molecules and the linker end-group, had been previously described mathematically by Elovich's and Polanyi-Wigner equations [38] and would lead to a simple first-order response of the adsorbed gas-molecule being released back into a gaseous form. Figure 13 shows the normalized conductance versus time plot of one complete cycle of analyte exposure to the device and its subsequent removal, curve-fitted with the Elovich equation for the adsorption cycle and Polanyi-Wigner equation for the desorption cycle. The plot suggests that the mathematical model was in decent agreement with the experimental data. The average root-mean-square-error of the curve-fitting was ~48% of the average data.

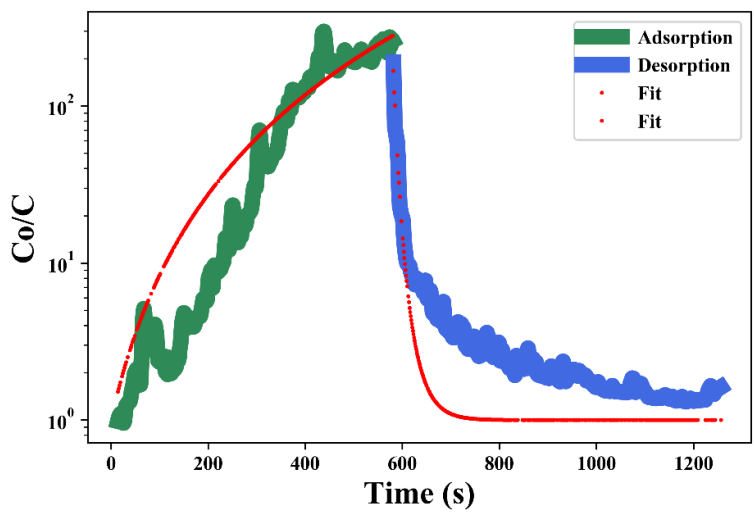

**Figure. 13** Normalized conductance response to one period of cadaverine exposure to nanogap device and subsequent removal curve-fitted to $2^{nd}$ order adsorption-desorption models.

## CONCLUSIONS

We present a new class of chemiresistors based on the capture of gas molecules within a nanoscale gap. We fabricated gas-sensing devices with gold electrodes separated by a ~6 nm nanogap functionalized with a fully conjugated ter-phenyl linker molecule, (4-((4-((4-mercaptophenyl)ethynyl)phenyl)ethynyl)benzoic acid) for electrostatic capture of cadaverine gas. We demonstrated ultra-low power resistance switching in batch-fabricated nanogap junctions upon detection of target analyte - cadaverine. The stand-by power consumption was measured to be less than 15.0 pW and the $R_{OFF}/R_{ON}$ ratio was more than eight orders of magnitude when exposed to ~80 ppm of cadaverine. A phenomenological electrical model of the device is also presented in good agreement with experimental observations. Cross-sensitivity of the gas sensor was tested by exposing the device to some of the commonly found VOCs and other atmospheric gases. The experiments revealed a highly selective sensor action against most of these analytes. These batch-fabricated sensors consume ultra-low power and demonstrate high selectivity; therefore, they can be suitable candidates for sensor applications in power-critical IoT applications and low power sensing.

**ACKNOWLEDGEMENTS**

The authors would like to acknowledge the contribution of Prattaydeepta Kairy, Chayanjit Ghosh and Navid Farhoudi for assisting with various aspects of the sensor testing. This work made use of University of Utah USTAR shared facilities support, in part, by the MRSEC Program of NSF under Award No. DMR-1121252. This work was sponsored under cooperative agreement HR0011-15-2-0049 of the DARPA N-ZERO program.